# Authorship Without Writing: Large Language Models and the "Senior Author" Analogy

Clint Hurshman, Sebastian Porsdam Mann,
Julian Savulescu, & Brian D. Earp

**Authors and Affiliations:**
Clint Hurshman[1] (clinthurshman@nus.edu.sg), Sebastian Porsdam Mann[1,2] (sebastian.porsdam.mann@jur.ku.dk), Julian Savulescu[1,3] (jsavules@nus.edu.sg), and Brian D. Earp[1] (bdearp@nus.edu.sg)
[1]Centre for Biomedical Ethics, Yong Loo Lin School of Medicine, National University of Singapore, Singapore
[2]Centre for Advanced Studies in Bioscience Innovation Law, University of Copenhagen, Copenhagen, Denmark
[3]Uehiro Oxford Institute, University of Oxford, Oxford, United Kingdom

**Abstract:** The use of large language models (LLMs) in bioethical, scientific, and medical writing remains controversial. While there is broad agreement in some circles that LLMs cannot count as authors, there is no consensus about whether and how humans using LLMs can count as authors. In many fields, authorship is distributed among large teams of researchers, some of whom—including paradigmatic "senior authors" who guide and determine the scope of a project and ultimately vouch for its integrity—may not *write* a single word. In this paper, we argue that LLM use (under specific conditions) is analogous to a form of senior authorship. On this view, the use of LLMs, even to generate complete drafts of research papers, can be considered a legitimate form of authorship according to the accepted criteria in many fields. We conclude that either such use should be recognized as legitimate, or current criteria for authorship require fundamental revision. **AI use declaration:** Chat GPT version 5 was used to help format Box 1. AI was not used for any other part of the preparation or writing of this manuscript.

**Keywords:**

Research ethics, professional ethics, philosophy, biomedical research\

# Authorship Without Writing:

# Large Language Models and the "Senior Author" Analogy

Clint Hurshman, Sebastian Porsdam Mann, Julian Savulescu, Brian D. Earp

## I. Introduction

The use of large language models (LLMs) in bioethics as well as scientific and medical writing continues to be controversial. Thus far, there has been broad agreement—for example, among medical publishers—that *LLMs* cannot count as authors, but there is still no consensus about the status of LLM-assisted text production as a form of writing, and by extension, the status of LLM *users* as authors. Here, we contribute to this debate by exploring—and drawing analogies to—the collaborative nature of writing, and the distributed character of authorship, in many domains of research.

Even before the popularization of LLMs, established norms of authorship (as exemplified by widely accepted authorship criteria set forth by the International Committee of Medical Journal Editors) allowed that some members of research teams could count as authors, *even without writing a single word*—provided they contribute to the work in other, sufficiently substantive ways (e.g., data analysis). And while "gift" authorship or authorship on the basis of mere funding acquisition—or otherwise absent a meaningful intellectual contribution—is generally agreed to be unethical, other, more substantive forms of so-called "senior" authorship are accepted in many disciplines.

In this paper, we explore an analogy between this form of authorship and the use of LLMs to generate research manuscripts. We conceptualize LLM use (provided that other conditions are met, such as ideation, supervision, critical engagement with manuscript content, vetting, and so on) as a kind of *minimal senior authorship*: "minimal" in the sense that the LLM user may not write a single word directly in the manuscript (although, as we will see, this does not imply that she performs minimal work); and "senior" in the sense that the user's interaction with the LLM is analogous (in the ways relevant to authorship) to some senior authors' interactions with junior authors. Like those senior authors, an LLM user can achieve authorship without writing.[1] The upshot is that consistency with the prevailing authorship norms in bioethics as well as many medical and scientific fields supports treating LLM use by human researchers, even to generate complete drafts of research papers—and in limit cases, *without any further drafting or editing by the human user*—as a form of authorship in some cases. We conclude that either this form of authorship should indeed be recognized as legitimate, or the prevailing authorship norms in many fields should be revised, thus likely resulting in the loss of authorship status for many human researchers who currently make substantial contributions to jointly authored pieces.

The structure of the paper is as follows. In Section II, we present two case studies to illustrate the analogy between senior authorship and the use of an LLM to generate research manuscripts. In Section III, we develop a simple argument for the claim that Jones is an author of the paper she submits. In Section IV, we address a series of objections to this simple argument. Section V concludes.

## II. Two cases

Consider the following two cases:

---

[1] Note that this does not imply "junior authorship" status on the part of LLMs; here, we remain agnostic about this separate issue. Note also that, in order to achieve this form of authorship, it is not at all necessary that the author is literally a *senior* researcher (i.e. a tenured professor, as opposed to a graduate student); rather, junior researchers, too, can achieve something analogous to senior authorship by guiding an LLM and critically engaging with its outputs.

*Junior Author*. Professor Smith is a Principal Investigator (PI) on a large grant who employs a junior researcher to help her generate research papers. She instructs her postdoc, Charlie, to write a draft of a paper, providing him with some key parameters including the basic idea for the thesis, a rough structure for the intended argument, and some representative citations. After Charlie produces the draft, Smith carefully evaluates the text and makes various critical suggestions for improvement. Charlie then makes the requested changes. Smith again reviews the manuscript to confirm argument quality, citation accuracy, and that the paper makes a contribution to the field. Satisfied that it does, she adds her name to the paper as senior author. Smith and Charlie explicitly disclose their respective roles in producing the manuscript (i.e., by making a formal CRediT[2] statement), and the two submit the paper to a bioethics journal.

*LLM*. Professor Jones is a Principal Investigator (PI) on a large grant who employs an LLM called "Charlie" to help her generate research papers. She instructs Charlie to write a draft of a paper, providing the LLM with some key parameters including the basic idea for the thesis, a rough structure for the intended argument, and some representative citations. After Charlie produces the draft, Jones carefully evaluates the text and makes various critical suggestions for improvement. Charlie then makes the requested changes. Jones again reviews the manuscript to confirm argument quality, citation accuracy, and that the paper makes a contribution to the field. Satisfied that it does, she adds her name to the paper as "senior" (and sole) author. Jones explicitly discloses her role in producing the manuscript (i.e., by making a formal CRediT statement), further adds an AI use statement acknowledging Charlie's contribution, and submits the paper to a bioethics journal.

*Junior Author* describes common practice in academia. In many scientific fields, authorship is *distributed*. In 2021, a paper in *Anaesthesia* on COVID-19 vaccine modeling was published with over 15,000 coauthors (COVIDSurg Collaborative and GlobalSurg Collaborative 2021). While this case is unusual—it currently holds the record for most authors on a published research article—it illustrates the radical division of labor among "authors" who play various roles in scientific inquiry such as designing and executing experiments, collecting and interpreting data, reviewing other published research, and of course writing and editing text. "Authorship" has become a technical notion, used to acknowledge participants' contributions to a work of research, although some of them never actually put pen to paper.

---

[2] The Contributor Role Taxonomy (CRediT) provides a set of fourteen roles, and is used in cases such as those above to make the contributions of members of a research team more transparent (CRediT - Contributor Role Taxonomy n.d.).

In our own fields, philosophical and empirical bioethics, work is highly interdisciplinary, with a variety of models of collaboration. Borry and colleagues (2006; cited in Resnick and Master 2014) found that empirical papers in bioethics had 2.97 authors on average, while "conceptual" papers had 1.35. In philosophy—as in most other humanities disciplines—coauthorship was historically uncommon, due both to institutional pressures and a less-clear division of labor than in, say, laboratory environments. However, even here, coauthoring is becoming increasingly prevalent. Analyzing data from the online database PhilPapers, David Bourget and Justin Weinberg (2021) found that the percentage of papers in major philosophy journals with multiple authors had risen from 5% in 1900 to 17% in 2021.

While *Junior Author* is intentionally contrived, many of these coauthored projects share its basic structure: a single (typically junior) author writes a draft, incorporating ideas from their discussions with the senior author(s), the latter of whom may have initially brainstormed the idea based on their knowledge of the field and what is likely to count as a meaningful contribution to the literature. The senior author(s) will later provide feedback to refine the arguments and correct any errors they notice, and may—or may not, depending on the quality of the initial draft and the junior author's skill level and ability to take direction—tweak or add to the text themselves. Finally, the team submits the article for publication, with each of their names on it. Insofar as this common practice is justified, Smith is a genuine author of her research paper.[3] Indeed, she clearly meets the widely accepted criteria for authorship set out by the International Committee of Medical Journal Editors (ICMJE), as shown in Box 1.

> Box 1. How Smith meets the ICMJE criteria for authorship
>
> The ICMJE requires that an author meet all four of the following criteria (2019 version):
>
> 1. **Substantial contributions** to the conception/design of the work *or* acquisition/analysis/interpretation of data.
> 2. **Drafting the work or revising it critically** for important intellectual content.
> 3. **Final approval** of the version to be published.
> 4. **Accountability** for all aspects of the work, ensuring integrity and accuracy.
>
> Applying these to Professor Smith:
>
> 1. **Criterion 1 (Conception/design):**
>    ✓ Smith contributes the **basic idea for the thesis, rough structure, and key citations**. That counts as substantial input into conception/design.
> 2. **Criterion 2 (Drafting/revising):**
>    ✓ While Charlie produces the draft, Smith engages in **critical review and suggestions for improvement**—which is explicitly recognized by ICMJE as satisfying the "revising it critically" clause.
> 3. **Criterion 3 (Final approval):**
>    ✓ Smith **reviews and signs off** on the manuscript after confirming its quality and contribution.
> 4. **Criterion 4 (Accountability):**
>    ✓ By attaching her name, she accepts **responsibility for the integrity** of the work.

The norms surrounding research produced with the help of generative artificial intelligence (AI) are less clear in comparison (Ganjavi et al. 2024, An et al. 2025). Since the release of ChatGPT in November 2022, researchers have incorporated it (and other LLMs) into many stages of the research process, including the review of literature, generation of hypotheses and arguments—and, of course, the drafting of text. However, these uses continue to be controversial, due both to concerns about the reliability of LLMs, and to lack of clarity about how to allocate credit for their outputs (Porsdam Mann et al. 2023b, Khosrowi et al. 2024, Formosa et al. 2025). Should LLMs themselves be credited as authors (for discussion, see Levy 2025)? Should LLM *users* be credited as authors just as though they had written every word themselves? If LLM use should be disclosed without crediting it as an author, *which* uses must be disclosed (Resnick and Hosseini 2025)? And so on.

---

[3] Contribution is generally acknowledged in descending author order, with the exception that the last or senior author may have contributed significantly to the design, conceptualization and/or funding of the work. Joint first or joint last authorship is becoming more common to acknowledge contributions to these key positions which are often seen as prestigious.

While several research articles, published soon after the release of ChatGPT, listed it as an author (see Stokel-Walker 2023), many journals have since banned that practice. And yet, despite some widespread hostility to LLMs in general, few academic journals ban its use outright. As Hosseini and colleagues (2023) argue, bans would be difficult to enforce and would encourage authors to hide their use. Instead, many journals simply require authors to disclose their use of LLMs. On the other hand, some journals, such as *The New England Journal of Medicine AI*, now even *encourage* the use of AI in submissions (Koller et al. 2024).

Three philosophical questions—two conceptual, and one ethical—about LLM use should be separated:

1. Can human researchers who prompt and edit (or prompt the LLM to edit) outputs of LLMs legitimately count as authors of those outputs?
2. Can LLMs ever legitimately count as authors of their own outputs?
3. In what ways, if any, should human researchers use LLMs to write research papers?

Regarding the third question, the use of LLMs raises a variety of ethical concerns, not only about distribution of credit (Earp et al. 2024, Khan et al. 2025) but also about potential deskilling or even wholesale replacement of human labor by generative AI in certain domains, which is increasingly occurring in creative, customer-service, and programming work (Klarna 2024, February 27; AbuMusab 2024, Brynjolfsson et al. 2025). There are also pedagogical concerns—for example, that students will not learn how to write on their own, or even how to think, since "writing is thinking" (Editor 2025; see also Lemasters and Hurshman 2024, Dabbagh et al. 2025, Rodger et al. 2025, Sparrow and Flenady 2025). Such use even raises concerns about (*inter alia*) sustainability, given the high energy and water costs of training and running generative AI models (Google 2025, Li et al. 2025). While these issues are important, we do not focus on them in the present paper. We will also set aside the second question, though it reappears below in Section IV.4. This paper will primarily focus on the first question.

The aforementioned controversies notwithstanding, there is, we suggest, a simple analogy to be drawn between Smith (the senior author) and Jones (the LLM user). Each of them guides, or "prompts," an assistant to produce a text based on an idea they had, provides some key parameters for how the argument should go, and critically reviews the draft as necessary to ensure its integrity and/or to further shape its content at a more fine-grained level. We may even suppose that each of them provides the *same* parameters, receives the *same* draft, and provides the *same* feedback. If their contributions are identical in these ways, then it seems that they should be considered just as *authorial*—see Box 1 for details—as well. Arguably, the only difference between the two cases lies in the nature of the assistant or *collaborating entity*, not with Jones or Smith themselves. Therefore, we suggest that consistency with the authorship practices illustrated in *Junior Author* implies that Jones in *LLM* is a genuine author of her paper as well.

### III. The argument

Our argument is as follows:

1. Smith is an author.

2. If Smith is an author, then Jones is an author.

3. Therefore, Jones is an author.

The upshot, then, is that it is possible to author an article with the assistance of an LLM, as Jones does. Note that this is not to say that Jones *coauthors*, since it does not follow that the LLM itself is an author. Despite this, Jones' role is that functional equivalent of that of a *senior* author like Smith. Insofar as their roles are equivalent, they seem, also, to be equally authorial. The conclusion, we argue, follows from the analogy between LLM prompting and the practices of senior authorship that are widely accepted in research fields. Consider each of the premises in turn.

*III.1 Premise 1*

The first premise is that Smith is an author, despite the fact that she does not (directly) write a single word of her manuscript. As noted above, this is not unusual. Although there is some degree of vagueness in how authorship is understood in various fields, the widely accepted ICMJE criteria (listed in Box 1) provide some crucial guidance. Specifically, the ICMJE recommends that authorship be based on four criteria, *all four* of which must be met by each individual who can legitimately be counted as an author. As a reminder, these are:

- Substantial contributions to the conception or design of the work; or the acquisition, analysis, or interpretation of data for the work; AND

- Drafting the work or reviewing it critically for important intellectual content; AND

- Final approval of the version to be published; AND

- Agreement to be accountable for all aspects of the work in ensuring that questions related to the accuracy or integrity of any part of the work are appropriately investigated and resolved.

The ICMJE further explains that "[c]ontributors who meet fewer than all 4 of the above criteria for authorship should not be listed as authors, but they should be acknowledged."

So, meeting these four criteria is necessary and, we take it, *sufficient* for a person's contribution to rise to the level of (co)authorship.[4] And as noted in Box 1, Smith meets all of them: she contributes to the conception and design of the work by providing Charlie with the thesis and structure that he develops into the full draft; while she doesn't draft the work herself, she reviews it critically and provides feedback for Charlie to incorporate; she approves the final version and, by submitting it, agrees to be held accountable for the content therein. Therefore, according to the authorship norms of medical research, bioethics, and many other disciplines, Smith is clearly an author.[5]

### III.2 Premise 2

We take it that the connection between *Junior Author* and *LLM* is clear. Smith and Jones appear to contribute to their respective research papers in the same ways (though we consider possible disanalogies in Section III.2). If their contributions are identical in all relevant respects, then, insofar as authorship status is determined by one's contributions to a paper, so must be their status as authors. If Smith is an author, then so is Jones. To deny authorship to Jones would require denying it to Smith as well, which is implausible—that is, *assuming* one accepts the ICMJE or other similar criteria, a point we'll return to—for the reasons noted above.

---

[4] The *sufficiency* of these criteria is important for several reasons. First, treating the conditions as *sufficient* for authorship mitigates the risk that contributors to a project are unfairly denied credit. Second, contributors may be strategically omitted from a list of authors in order to conceal conflicts of interest. For these reasons, we take it that the ICMJE's guidelines are intended as *sufficient* conditions as well as necessary ones.

[5] Medical research is representative, in this respect, of many fields, particularly in the sciences (see Winsberg et al. 2014). See section IV.1 below.

### *III.3 Conclusion*

By *modus ponens*, Jones is an author. The mere fact that an LLM assisted her, even to the extent that she did not contribute any words directly to her manuscript, does not by itself undermine her status as an author, since her contribution to the manuscript content is exactly analogous to that of a "senior" author who supervises, critiques, and vouches for the writing of a junior author.

Our claim is primarily conceptual, and normative only in an indirect sense. That is, again, our aim is not primarily to say whether authors *should* use LLMs to write papers, only that doing so can still be understood as a form of authorship. This claim is compatible with there being good reasons not to write with LLMs, and indeed for journals and institutions to discourage such practices; these normative considerations are a separate issue.

If there are such reasons, however, we stress that they do not follow from a deficiency of the senior author's contribution *qua* authorial contribution. On the other hand, our argument (particularly premise 1) is not completely value-neutral, insofar as it takes existing authorship norms, as exemplified by the ICMJE guidelines, for granted. Authorship norms are contested, so one could fairly argue that these guidelines should be revised, but this, too, is a separate issue. Our conclusion is therefore partly an ameliorative one: that Jones has *the kind of authorship* that is (taken to be) at issue in scholarly contexts; or, put differently, the *consistent* application of norms for senior authorship would cover Jones's contribution as well as Smith's.

### IV. Objections and replies

We will now address a series of objections that might be levied against the foregoing argument. First, we consider two objections that each deny one of the premises of the argument. We then consider a practical objection concerning the difficulty of ensuring the reliability of LLM outputs, and an objection that holds that LLMs cannot be authors because they lack some essential features of authorship, such as agency or sentience.

*IV.1 Denying Premise 1*

Since Smith does not contribute a single word to Charlie's draft, it may sound odd to some readers to call her an author. If she is not an author, then it is consistent to deny that Jones is an author, too. Indeed, this move is attractive because the ICMJE guidelines above are vague. What, after all, counts as a "substantial contribution"? Is it unreasonable to treat the direct composition of at least some text as a necessary condition for substantial contribution? Partly because of this vagueness, the criteria can be abused (e.g., strategically including or omitting people from the list of authors) and can have the effect of flattening the contributions of coauthors who may contribute to projects in importantly different ways. For these reasons, again, authorship norms are contested, and we are not opposed in principle to their being revised.

In response, however, we again point out that "authorship" in research is a technical notion that does not necessarily track the concept of authorship as it is used in other domains or among laypeople. For research purposes, a "vague" notion of authorship is needed in order to acknowledge the heterogeneous contributions of (especially junior) researchers that may be difficult or impossible to capture under a more fine-grained definition. Certainly, a requirement that all authors put words to paper would be too strong. Applied to the COVIDSurg - GlobalSurg (2021) paper, for example, this requirement would obscure the contributions of thousands of people around the world who contributed to the collection and interpretation of data. So, there are good *prima facie* reasons to maintain a concept of authorship that is permissive enough to apply to Smith.

Crucially, in other domains, different concepts of authorship may be appropriate. In the production of literary prose and poetry, for example, the articulation of ideas and images is one of the most important forms of work that the author performs; there, the use of an LLM might conflict with authorship in a more fundamental way. Similarly, in law, every "author" on a paper is generally assumed to have contributed some words directly to the manuscript, and to be accountable for the content therein (Tietz and Price 2020). Perhaps for the same reason, however, *coauthoring* in such domains is relatively uncommon and, when it does occur, it often calls for a more fine-grained explication of the division of labor.

In short, authorship norms are contextually variant and subject to revision, but for research purposes there are strong reasons to maintain a concept of authorship consistent with premise 1.

## *IV.2 Denying Premise 2*

Alternatively, one could deny premise 2 by insisting that there is a disanalogy between the contributions of Smith and Jones to their respective research papers. Most obviously, while Jones's authorship is fundamentally a *solitary* process (i.e., on the assumption that LLMs are not sentient agents and do not provide the same kind of collaboration or companionship as humans do), Smith's collaboration with Charlie occurs in the context of a social relationship: it is a *coauthorship*. Smith is a *mentor* to Charlie, and if she used an LLM instead, then Charlie would miss out on the goods of mentorship. Indeed, if LLM-assisted authorship becomes widely practiced, then opportunities may disappear for junior researchers to contribute to coauthored papers, or even to have jobs. We are sympathetic to this worry, as well, but we have two responses.

First, our argument only concerns the extension of the term "authorship," not how it should ethically be practiced. So, our view is compatible with structural efforts by institutions and journals to protect opportunities for junior researchers, which may include discouraging LLM use. Our argument only implies that such measures should *not* be justified on the grounds that authorship assisted by an LLM is not authorship. Rather, the justification might be that the social goods (such as mentorship) that human collaboration produces outweigh the benefits of allowing LLM-assisted authorship, the latter's legitimacy *qua* authorship notwithstanding.

Second, as far as social goods are concerned, we should not be too dismissive of the potential benefits of LLM-assisted authorship. For example, researchers at under-resourced institutions are often unable to hire junior researchers with whom to collaborate. These researchers might benefit from having the option to work with an artificial collaborator, in the same way that some non-native English speaking researchers benefit from using LLMs to help communicate with the English-speaking research community (Liao et al. 2024, cf. İlhan et al. 2024). Ideally, institutions should aim to adopt policies that allow researchers to reap these benefits while also protecting the social goods associated with human collaboration.

In short, then, while there are significant disanalogies between what Smith and Jones each do (most notably, in the social goods they produce), these are not disanalogies in their authorial contributions as such. Moreover, they are complex: there are valuable social goods to be gained by either approach.

*IV.3 Responsibility*

Currently, LLMs often "hallucinate," or apparently fabricate, purported facts and citations when used to produce academic prose. An (albeit somewhat outdated) study by Bhattacharyya and colleagues (2023) found that when used to produce medical papers, ChatGPT only cited references correctly 7% of the time; 46% of references were real but somehow inaccurately cited, and the other 47% were non-existent. While the introduction of 'reasoning models' that simulate the kind of 'chain of thought' approach humans take to problem solving, as well as tool use such as web search, can be used to substantially reduce the incidence of hallucination, these technical solutions do not remove the risk entirely. Even highly sophisticated applications of LLMs in ideal, bounded settings (e.g., clinical note taking) continue to show a hallucination rate of 1-2%, though this is comparable to the error rate of humans (Asgari et al. 2025). As a result, one may question the *prudence* of relying on LLMs.

It goes without saying that researchers have a responsibility to vet the outputs of LLMs when relying on them to produce text for research papers (Porsdam Mann et al. 2023a). Failure to do so while still claiming authorship is dishonest and even dangerous because it could lead to the spread of misinformation, and indeed the reliance on bad information in high-stakes (e.g. legal, medical, scientific) contexts. Moreover, the ICMJE guidelines point out, "reviewing" a work "critically for important intellectual content" is a constitutive, that is, necessary feature of authorship, particularly when one isn't involved in the initial drafting of the piece. So, some (irresponsible) LLM users do not count as authors, but this is no reason to doubt that a senior author who *does* review the work critically thereby counts as one. Similarly, Smith is required to check the referencing and work of the human research assistant, who might also make mistakes, and take responsibility for it.

We motivated premise 2 by arguing that Smith's and Jones's contributions to their respective articles are directly analogous. However, in real cases, this may not strictly be true.

Given the limitations of LLMs, in order for Jones to do due diligence, she likely needs to engage *more* critically with the LLM manuscript than does Smith. Charlie the human can, one hopes, be trusted not to make stupid mistakes, but "Charlie" the LLM cannot. Thus, in order for Jones to be as confident as Smith in the quality of her manuscript, and so be able to *responsibly* take responsibility for it, she likely faces a higher burden.

However, whether such responsibility-taking is a *necessary* condition for authorship has been contested. Levy (2025) argues that the expectation, as identified in ICMJE guidelines, that authors each individually be accountable for the entire content of their papers is too strong for fields like medical research, where labor is often so radically distributed that no individual author—not even a senior author!—can be wholly accountable for ensuring the integrity of *every* aspect of a work (see also Winsberg et al. 2014). Even so, one might argue that *some* human author must be accountable for each aspect of a work, and that a human, perhaps a senior author, must ultimately be accountable for ensuring that *other* human collaborators are sufficiently trustworthy to be relied upon in taking responsibility for *their* contributions, even if the senior author cannot, themselves, personally vouch for every feature of a manuscript (e.g., confirming that a complex statistical analysis was performed properly).

In the case of a paper produced by a single human and an LLM, these more elaborate criteria might not be met unless the human can, in fact, vouch for *every* part of the paper—even the parts produced by an LLM. In other words, since the LLM cannot be accountable for its contribution to the manuscript, it may be that Jones in our example is obligated to *assume* responsibility (see Lang et al. 2023). Paradoxically, however, this suggests that if Jones uses the LLM responsibly, making an adequate effort to ensure the integrity of her work, then her contribution may in some sense be *more* authorial than that of Smith.

At this point, it could be further argued that the widespread use of LLMs to assist research will *inevitably* (or very likely) lead to irresponsible use, and that LLM-assisted authorship should be opposed on this basis. However, we have already conceded that there may be good ethical reasons to oppose LLM-assisted authorship without this entailing that such authorship is not authorship (properly so called). Nevertheless, let us pursue this objection to see what may be learned. Why might one expect that greater acceptance of LLM-assisted authorship would likely or even inevitably lead to irresponsible use of LLMs in the production of academic work?

This could occur for at least three reasons. First, researchers in academia are already under intense pressure to "publish or perish." They therefore have incentives to produce and submit articles as quickly as possible, which sometimes leads to the submission of poor-quality work. This pressure to publish has been cited as one of the causes of the "replication crisis" in psychology (Earp and Trafimow 2015; Everett and Earp 2015), and one could reasonably worry that it will lead some authors to submit LLM-produced work without adequately checking its integrity. (Indeed, this is already occurring, but one may worry that it will be exacerbated if LLM-assisted writing becomes more prevalent or accepted.)

Second, one might worry that the cognitive effects of relying exclusively—or even predominantly or significantly—on AI outputs rather than drafting substantial portions of a paper "by hand" (see Kosmyna et al. 2025) will lead not only to the atrophying of users' own ability to write, but also their ability to critically engage with LLM-produced texts to ensure their integrity (see Voinea et al. in preparation). Our tendency to depend on cognitive scaffolds when doing so is reasonably effective can easily lead to overreliance, though this may be unintended and even unrecognized. We concede this is empirically plausible. However, early meta-analytic evidence at the college level suggests that the effects of LLM assistance depends on *how* they are used;

when used in structured, pedagogically informed ways, such as within problem-based learning frameworks or acting as an intelligent tutor, they can be associated with substantial gains in learning performance and moderate improvements in critical and creative thinking (Wang & Fan, 2025).

The third reason for which it might be expected that the prevalence of LLM-assisted authorship will very likely lead to irresponsible LLM use has to do with the *accessibility* of participation in scholarly research. As noted above, LLMs could increase access to a form of senior authorship for researchers who lack access to human collaborators like students or postdocs. Presumably, these aspiring senior authors will need the requisite skills to engage critically with LLM-produced texts and ensure their integrity. However, it is also possible that researchers who lack those skills, such as students who are still learning to research responsibly, will see LLM-assisted authorship as a way to contribute to scholarly discussions, and thus (even accidentally) submit inadequately vetted work.

It is, of course, an empirical question to what extent the acceptance of LLM-assisted authorship will actually lead, by these mechanisms or others, to a proliferation of bad research articles, and it's beyond the scope of the present paper to anticipate the answers to such empirical questions. We must reiterate, however, that the *practical* question of how to minimize the risk of publishing irresponsibly reviewed, LLM-produced manuscripts is separate from the *conceptual* question of whether a researcher who does responsibly review such a manuscript thereby counts as an author. So, again, efforts to mitigate the aforementioned effects should not be justified on the basis that the human authors of such papers are not really authors.

*IV.4 Moral Agency and Sentience*

Finally, it may be argued that authorship, like communication in general, is an essentially *intentional* act—an author is someone who consciously means to communicate something through the written word. Since the proximal mechanism by which text is produced in the Jones example, namely by next-token prediction by an LLM, does not itself involve any such an intention, perhaps the writing that is produced has no author (Ostertag 2023). Let us see how this objection might go.

According to Grice's (1989) influential theory, communication involves not merely the production of signs, but the intention to make one's meaning understood through the use of those signs, and to be recognized as acting with such an intention. Communication, then, requires a degree of agency and perhaps even sentience which presently existing LLMs seem to lack, and which it is unclear whether future LLMs will ever possess. Since it is an LLM in the Jones example that produces the relevant signs, these signs—according to this argument—are unauthored.

Note, however, that the conclusion of our argument concerns the authorial status of *Jones*, not of the LLM itself. It may be that the LLM is not comparable to a junior author in important respects, such as by having agency and intention in drafting material, though the "senior author" retains the essential features of authorship. For example, it could be argued that the larger functional system that includes both the human and the LLM does have intention (by virtue of the human participant), such that by intentionally prompting the LLM, reviewing its output, and—so to speak— "assenting" to the output as an expression of what she wants to say, Jones thereby *imbues* the signs produced by the LLM with her intention and authorial authority.

On this interpretation, Jones's contribution is probably less authorial than that of a solo author who conceives and writes a paper entirely by herself—and so, to avoid misleading readers, she should disclose her use of the LLM (Verhoeven et al. 2023). But the relevant comparison here is not to a "traditional" or (as one might say) "artesanal" author who writes each word of a manuscript herself, but rather, to *Smith*, who meets current authorship criteria despite also not composing the text of her manuscript directly. Our contention is that the contributions of Jones, notwithstanding her LLM use, are no less authorial than those of Smith.

The analysis just given assumes that LLMs—perhaps by virtue of their lack of agency or intention—cannot make an authorial contribution. Rather, it is the human on our analysis who "imbues" the work with her own intention to communicate by endorsing it as an expression of the same. However, perhaps we should not be so quick to rule out an authorial contribution by the LLM. After all, the "authoriality" of an LLM's output can be understood from either an *internal* or an *external* perspective. And while LLMs probably do not count as authors based on internal criteria, we should be open to the possibility of their counting as such based on external criteria.

Let us explain. We are happy to concede, in line with the objection we are entertaining in this section, that LLMs lack *intrinsic* features like intentionality that are generally seen as necessary for an entity to deserve credit or blame, or to be morally responsible for text it produces.[6] Insofar as LLMs lack these capacities, they are incapable of being authors in the internalist sense.

On the other hand, from an "externalist" perspective, LLMs have many of the features of genuine coauthors. In other words, Jones' LLM has a "system function" (Cummins 1975) comparable to that of a human coauthor. By analogy, some of the authors of this paper have argued that LLMs can play the role of identifying relevant reasons in an argument which can then be evaluated for their cogency, relevance, and so on, by human decision-makers (Porsdam Mann et al. 2024). LLMs differ from human interlocutors in that they cannot be responsible for those reasons, and may not even understand them as reasons, for example, by subjectively "grasping" their force (see Earp 2012). But for many purposes, at least, it is irrelevant how the reasons in a shared decision-making process are produced, so long as they constitute genuinely good reasons as can be confirmed by the human-in-the-loop.

---

[6] Because they lack interests, LLMs are also incapable of benefiting from credit for authorship (and of being harmed by being blamed for making mistakes). We think that this is a potentially important albeit underexplored disanalogy between LLMs and human authors, but we will not pursue it further in the present paper.

In fact, we think there is at least one somewhat weighty externalist reason to consider LLMs as full-fledged authors (in the technical sense in which that term is employed within academia). If LLMs cannot be authors, Jones in our example would be counted as both first and last (i.e., as the sole) author, whereas Smith, who did the same amount of work, would only be *last* author while Charlie, her postdoc, would be the first author. And that might seem rather unfair.

Author position designates relative contribution and is currently seen as important in assigning responsibility, credit, recognition, promotion and other academic goods (cf. Koplin 2023). Thus, even if Jones accurately acknowledges the contribution of the LLM as stated in our example, the "record" would show Jones to be first—and only—author, whereas Smith's authorship status would be diluted. Assuming that this current paradigm for interpreting authorship is justified, or at any rate, that it is likely to persist and have real-world consequences for authors, this gives us a pragmatic reason to place the LLM in the appropriate authorial position. Nevertheless it is beyond the scope of the present paper to commit to a specific policy for disclosing and crediting LLM use, whether in relation to potential LLM authorship crediting/placement or otherwise.

### V.   Conclusion

We have argued that the use of LLMs even to generate entire papers to which one then makes minor revisions (e.g., through offering critical suggestions which the LLM then incorporates) can be consistent with a form of authorship, namely "minimal senior" authorship, that is already widely practiced in bioethics and similar fields. While the present paper has focused on the authorial status of LLM *users* in research contexts, there remain philosophical questions about the authorial status of LLMs and the ethics of using them, as well as policy questions about how LLM use should be disclosed and how credit should be allocated for works thereby produced. We have additionally argued that there are good pragmatic reasons to accord LLMs appropriate authorial credit in order to properly evaluate the contributions of human coauthors.

In conclusion, we suggest that either the form of authorship Jones exhibits should be recognized as legitimate (by *modus ponens*), *or* the prevailing authorship norms in many fields should be revised, thus likely resulting in the loss of authorship status for many human researchers who currently make substantial contributions to jointly authored pieces (by *modus tollens*).

Preprint. Paper under review.

**Funding:**

Clint Hurshman: CH's research for this paper is supported by NUSMed and ODPRT (NUHSRO/2024/035/Startup/04) for the project "Experimental Philosophical Bioethics and Relational Moral Psychology" with BDE as PI.

Sebastian Porsdam Mann: SPM's research for this paper was supported by a Novo Nordisk Foundation Grant for a scientifically independent International Collaborative Bioscience Innovation & Law Programme (Inter-CeBIL programme - grant no. NNF23SA0087056).

Julian Savulescu: This research is supported by the National Research Foundation, Singapore under its AI Singapore Programme (AISG Award No: AISG3-GV-2023-012)
This research project is supported by National University of Singapore under the NUS Start-Up grant; NUHSRO/2022/078/Startup/13)

Brian D. Earp: This research is supported by NUSMed and ODPRT (NUHSRO/2024/035/Startup/04) for the project "Experimental Philosophical Bioethics and Relational Moral Psychology" with BDE as PI. This research is supported by the National Research Foundation, Singapore under its AI Singapore Programme (AISG Award No: AISG3-GV-2023-012). This research is supported by a Google DeepMind Level 2 GIG (Googler Initiated Grant), "Understanding Norms for Human-AI Relationships," awarded to BDE as PI.

**Conflict of interest:** JS and SPM are advisors of AminoChain, Inc. JS is a Bioethics Committee consultant for Bayer and a Bioethics Advisor to the Hevolution Foundation.